# CCD POLARIMETRY OF DISTANT COMETS C/2010 S1 (LINEAR) AND C/2010 R1 (LINEAR) AT THE 6-M TELESCOPE OF THE SAO RAS

Oleksandra V. Ivanova[a*], Janna M. Dlugach[a], Viktor L. Afanasiev[b], Volodymyr M. Reshetnyk[a,c], Pavlo P. Korsun[a]

[a] *Main Astronomical Observatory of National Academy of Sciences, Kyiv, Ukraine.*
[*] *Corresponding Author. E-mail address: sandra@mao.kiev.ua*
[b] *Special Astrophysical Observatory, Russian Academy of Sciences, Nizhnii Arkhyz, Russia.*
[c] *Taras Shevchenko National University, Kyiv, Ukraine.*

# ABSTRACT

We present first measurements of the degree of linear polarization of distant comets C/2010 S1 (LINEAR) and C/2010 R1 (LINEAR) at heliocentric distances $r = 5.9 – 7.0$ AU. Observations were carried out with the SCORPIO-2 focal reducer at the 6-m telescope of the Special Astrophysical Observatory (Russia). Both comets showed considerable level of activity (significant dust comae and tails) beyond a zone where water ice sublimation is negligible (up to 5 AU). Significant spatial variations both in the intensity and polarization are found in both comets. The slope of radial profiles of intensity changes gradually with the distance from the photocenter: from − 0.7 near the nucleus up to about − 1.3 for larger distances (up to 100000 km). The variation in polarization profiles indicates the non uniformity in the polarization distribution over the coma. The polarization degree over the coma gradually increases (in absolute value) with increasing the photocentric distance from of about −1.9% up to −3% for comet C/2010 S1 (LINEAR), and from of about −2.5% up to −3.5% for comet C/2010 R1 (LINEAR). These polarization values are significantly higher than typical value of the whole coma polarization (∼−1.5%) for comets at heliocentric distances less than 5 AU. The obtained photometric and polarimetric data are compared with those derived early for other comets at smaller heliocentric distances. Numerical modeling of light scattering characteristics was performed for media composed of particles with different refractive index, shape, and size. The computations were made by using the superposition $T$-matrix method. We obtained that for comet C/2010 S1 (LINEAR), the dust in the form of aggregates of overall radius $R$ ~ 1.3 μm composed of $N = 1000$ spherical monomers with radius $a = 0.1$ μm, refractive index $m = 1.65 + i\ 0.05$, allows to obtain a satisfactory agreement between the results of polarimetric observations of comet C/2010 S1 and computations.

# 1. Introduction

Physical nature of comets is known mainly from the observations of bright comets close to the Earth and the Sun (typically 1−2 AU). A few of the Jupiter, Neptune family comets and Kuiper-Belt objects have been investigated with space experiments, while observations at large heliocentric distances (more than 5 AU) are still scarce and episodic, thereby not covering very important stage of the development of cometary activity. Moreover, such observations are very useful in order to get more information about the origin of comets as well as their relation to Kuiper-Belt Objects and Centaurs (Belskaya et al., 2008 and Stansberry et al., 2008).

Activity of distant comets cannot be explained in the frame of standard model, whereas sublimation of more volatiles than water ice such as CO, $CO_2$, $N_2$ can serve as a probable explanation of this phenomenon (Gronkowski and Smela, 1998). The results of observations of these comets not only demonstrate their high activity at large heliocentric distances. An important specific feature of this activity is its long-term nature, in contrast to the often observed outburst activity (Ivanova et al., 2011). This feature allows to discuss the physical processes associated with such activity in more detail. An analysis of the distant cometary activity usually focuses on finding proper sources of energy required to release particles from the cometary nucleus that, in turn, leads to the formation of the observed coma. In this context reliable information on the physical properties of the cometary dust is of great importance. Available observational data show the difference between the activity of distant and short-period comets (Epifani et al., 2007, 2008 and Korsun et al., 2008, 2010, 2014). It is possible that the nature of the dust differs also.

Observations of distant comets in different modes (photometry, spectroscopy, and polarimetry) allow to obtain a useful information about dust properties in the cometary coma at large heliocentric distances. So far, polarimetric measurements were used to study physical properties of the dust in comets approaching the Sun. Since the dust in distant comets may differ from that in comets close to the Sun (e.g. Kolokolova et al., 2007), the measurements of polarization of distant comets are very important for study of their physical properties and evolution. However, there are no comets studied with polarimetric techniques at heliocentric distances more than 5 AU.

The program of photometric and spectral studies of distant comets with a high level of activity beyond Jupiter's orbit was started at the 6-m BTA telescope of the Special Astrophysical Observatory of the RAS (SAO RAS, Russia) in 2006. The series of spectroscopic and photometric data for these comets were obtained (Korsun et al., 2006, 2008, 2014). So far a few spectra for distant comets (with $q > 5$ AU) were published. The spectra of well-known periodic comet 29P/Schwassmann–Wachmann 1 show the CN and ionic emissions at distances of ~6 AU from the Sun (Cochran et al., 1980; Cochran and Cochran, 1991, Korsun et al., 2008). Ionic emissions were also detected in the spectra of long period comet C/2002 VQ94 (LINEAR) at the distance of 8.36

AU from the Sun (Korsun et al., 2014). Nevertheless, no emission exceeding 3σ level was found in the spectra of comets C/2003 WT42 (LINEAR) (Korsun et al., 2010) and C/2006 S3 (LONEOS) (Rousselot et al., 2014). Also after spectra processing of comet C/2010 S1 (LINEAR) (Ivanova et al., 2014) and C/2010 R1 we concluded that only continuum was detected (without any emissions in the spectra of distant comets).

Therefore we have started a comprehensive program of polarimetric observations of active distant comets using the modified universal focal reducer SCORPIO-2 (Afanasiev and Moiseev, 2011) mounted in the prime focus of the 6-m BTA telescope of the SAO RAS.

In this paper, we present and analyze the first results of polarimetric observations of two distant comets C/2010 S1 (LINEAR) (hereafter C/2010 S1) and C/2010 R1 (LINEAR) (hereafter C/2010 R1). Comet C/2010 S1 was detected on September 21.36, 2010 as asteroidal object of around 18th magnitude at heliocentric distance $r = 8.85$ AU and geocentric distance $\Delta = 8.54$ AU. Follow-up observations showed that the comet had a bright coma and a tail. The comet has a perihelion distance of 5.89 AU and passed its on May, 2013, an orbit inclination $i = 125.3°$ and eccentricity $e = 1.0018997$ (Marsden, 2010).

Comet C/2010 R1 was discovered also as non active object of 21st magnitude on September 4.15, 2010, at heliocentric and geocentric distances equal to 7.15 and 6.66 AU, respectively. In follow-up observations performed at Magdalena Ridge 2.4-m reflector, it was found that this object had both coma and small tail. The comet reached perihelion at 5.6 AU in 2012 May, and had $i = 156.9°$ and $e = 1.36652$ (Marsden, 2010).

## 2. Observations and reduction

Observations of comets C/2010 S1 and C/2010 R1 were carried out at the 6-m BTA telescope (SAO RAS, Russia) with the focal reducer SCORPIO-2 (Afanasiev and Moiseev, 2011). Comet C/2010 S1 was observed through the Johnson's V filter in the polarimetric mode on November 25, 2011, when $r = 7.01$ AU and $\Delta = 6.52$ AU. The phase angle of the comet was $\alpha = 7.3°$. On November 12, 2012, the comet was observed through the g-sdss filter (λ=4640 Å, FWHM= 1262 Å) at $r = 6.05$ AU and $\Delta = 5.87$ AU, and the phase angle $\alpha = 9.4°$. Comet C/2010 R1 was observed through the r-sdss filter (λ=6122 Å, FWHM= 1149 Å) on February 6, 2013. The heliocentric and geocentric distances were 5.94 AU and 5.57 AU, respectively, and $\alpha = 9.2°$. Table 1 presents the log of observations, specifying the date of observations (the mid-cycle time), the name of comet, the heliocentric and geocentric distances, the phase angle, the position angle of the scattering plane, the filter, the total exposure and number of images, and the mode of observations.

**Table1.** Log of the observations.

| Date, UT | Object | $r$, AU | $\Delta$, AU | $\alpha$, deg | $PA$, deg | Filter | $T_{exp}$, s / $N^*$ | Mode |
|---|---|---|---|---|---|---|---|---|
| Nov. 25.63, 2011 | C/2010 S1 | 7.01 | 6.52 | 7.3 | 92.5 | V | 10/3 | Image |
| Nov. 25.71, 2011 | C/2010 S1 | 7.01 | 6.52 | 7.3 | 92.5 | V | 30/18 | Impol |
| Nov. 12.66, 2012 | C/2010 S1 | 6.05 | 5.87 | 9.4 | 70.9 | g-sdss | 10/3 | Image |
| Nov. 12.69, 2012 | C/2010 S1 | 6.05 | 5.87 | 9.4 | 70.9 | g-sdss | 30/20 | Imapol |
| Feb. 06.13, 2013 | C/2010 R1 | 5.94 | 5.57 | 9.2 | 285.4 | r- sdss | 60/3 | Image |
| Feb. 06.18, 2013 | C/2010 R1 | 5.94 | 5.57 | 9.2 | 285.4 | r- sdss | 30/16 | Imapol |

## 2.1. Instruments

An E2V-42-90 CCD chip of 2048 × 2048 pixels was used as a detector. A full field of view of the detector is 6.1'×6.1' with an image scale of 0.18"/pix. To increase the signal/noise ratio of the observed data binning of 2×2 was applied to the polarimetric images. For the measurements of polarization, an optical scheme consisting of rotating phase plates and a fixed polarization analyzer was selected. The two Wollaston prisms, designated WOLL-1 and WOLL-2, and a dichroic polarization filter (POLAROID) were used as the polarization analyzers. For our observations of the distant comets, we used two modes: dichroic polarization filter and Wollaston prism WOLL-1.

### 2.1.1. Polaroid

The dichroic polarization analyzer mounted in the focal reducer is intended for the measurements of linear polarization of extended objects. The analyzer can be set in three fixed positions by the angle 0° and ±60°. The intensities in three angles of the polaroid are $I(x, y)_0$, $I(x, y)_{-60}$ and $I(x, y)_{+60}$. The measured Stokes $Q'$ and $U'$ parameters in each point of the image with the coordinates $x, y$ are given by the relations:

$$\begin{cases} Q' = \dfrac{2 \cdot I(x, y)_0 - I(x, y)_{-60} - I(x, y)_{+60}}{I(x, y)_0 + I(x, y)_{-60} + I(x, y)_{+60}} \\ \\ U' = \dfrac{1}{\sqrt{3}} \dfrac{I(x, y)_{+60} - I(x, y)_{-60}}{I(x, y)_0 + I(x, y)_{-60} + I(x, y)_{+60}} \end{cases} \quad (1)$$

The true values of the Stokes $Q$ and $U$ parameters are as follows:

$$\begin{cases} Q/I = U'\cos(2\varphi) - Q'\sin(2\varphi) \\ \\ U/I = U'\sin(2\varphi) + Q'\cos(2\varphi) \end{cases} \quad (2),$$

where $\varphi$ is the orientation of the principal axes of the prism.

The final degree of linear polarization *P* and the position angle of the polarization plane *PA* are obtained from relations:

$$\begin{cases} P = \sqrt{Q^2 + U^2} \\ PA = \frac{1}{2}\arctan\left(\frac{U}{Q}\right) \end{cases} \quad (3)$$

### 2.1.2. Single Wollaston Prism

In this polarimetric mode of the instrument, an achromatic half-wave phase plate and a Wollaston prism are installed before a prism. The Wollaston prism, separating the beam into the ordinary o and extraordinary e rays by 5 deg (2′ on the celestial sphere), is inserted in the parallel beam (Afanasiev et al., 2012). We use the mask (high 2′ and diameter 4′), which cuts non-overlapping area. For fixed positions of the $\lambda/2$ phase plate 0°, 45°, 22°.5 и 67°.5, a series of pairs of spectra $I(\lambda)_o$ and $I(\lambda)_e$ in mutually perpendicular polarization planes is obtained at the exit of the focal reducer.

In general, when the orientation of the principal axes of the analyzer $\varphi$ and the vector direction maximum speed phase plate $\theta$ are arbitrary, measured up to the transformation of rotation values of the Stokes parameters $Q'$ and $U'$, the linear polarization is given by the relations:

$$\begin{aligned} U' &= 0.5\left(\frac{I(\lambda)_o - I(\lambda)_e}{I(\lambda)_o + I(\lambda)_e}\right)_{\theta=0°} - 0.5\left(\frac{I(\lambda)_o - I(\lambda)_e}{I(\lambda)_o + I(\lambda)_e}\right)_{\theta=45°} \\ Q' &= 0.5\left(\frac{I(\lambda)_o - I(\lambda)_e}{I(\lambda)_o + I(\lambda)_e}\right)_{\theta=22.5°} - 0.5\left(\frac{I(\lambda)_o - I(\lambda)_e}{I(\lambda)_o + I(\lambda)_e}\right)_{\theta=67.5°} \end{aligned} \quad (4)$$

For this mode, the degree of linear polarization *P* and the position angle *PA* of the plane of polarization were calculated from the Eqs. (2) and (3), respectively. Detailed description of the instrument and the reduction procedure are given by Afanasiev and Amirkhanyan (2012).

### 2.2. Data processing

Each individual image of the comet and flat field is bias corrected. Flat field, which was obtained for each angle of the phase plate or polaroid, was normalized to the average count in the center of image, it takes into account the artificial instrumental polarization. After that each image of the comet was flat field corrected.

Sky background was estimated using the procedure, which builds a histogram of counts in the image. The count corresponding to the maximum is the level of the sky background which was subtracted from the images.

Since measurements are differential, we combine the images of the comet using only central isophote, which is located closest to the maximum of brightness of the comet. The removal of the traces of cosmic ray particles is done at the final stage of reduction via the robust parameter estimates. The random polarization errors are estimated from the direct measurements using the so-called method of robust statistics (Maronna et al., 2006). We first compute the statistical moments in the observational data and, then we exploit an IDL code implementing the robust statistics method that searches for significant deviations from the average polarimetric response. In case such deviations are found, their contribution into the average polarization is being evaluated. If the given data point appears noticeably affecting the average polarimetric response, it is excluded from the consideration. Thus, we rid of the measurements with unreasonably high impact on the average polarization.

## 3. Results of observations

### 3.1. *Morphology of comets*

The both comets showed significant activity (bright comae and long tails) beyond the Jovian orbit, where sublimation of water ice can be neglected. Coma morphology of comet C/2010 S1 and its evolution with time is displayed in Fig. 1. On November 25, 2011 ($r = 7.01$ AU), the comet showed only elongated in the sunward – antisunward direction dust coma with short tail (Fig. 1a). Extensive dusty coma with highly condensed material surrounding the nucleus and extended dust tail (Fig. 1b) were detected when the comet approached the Sun, on November 12, 2012, when $r = 6.05$ AU.

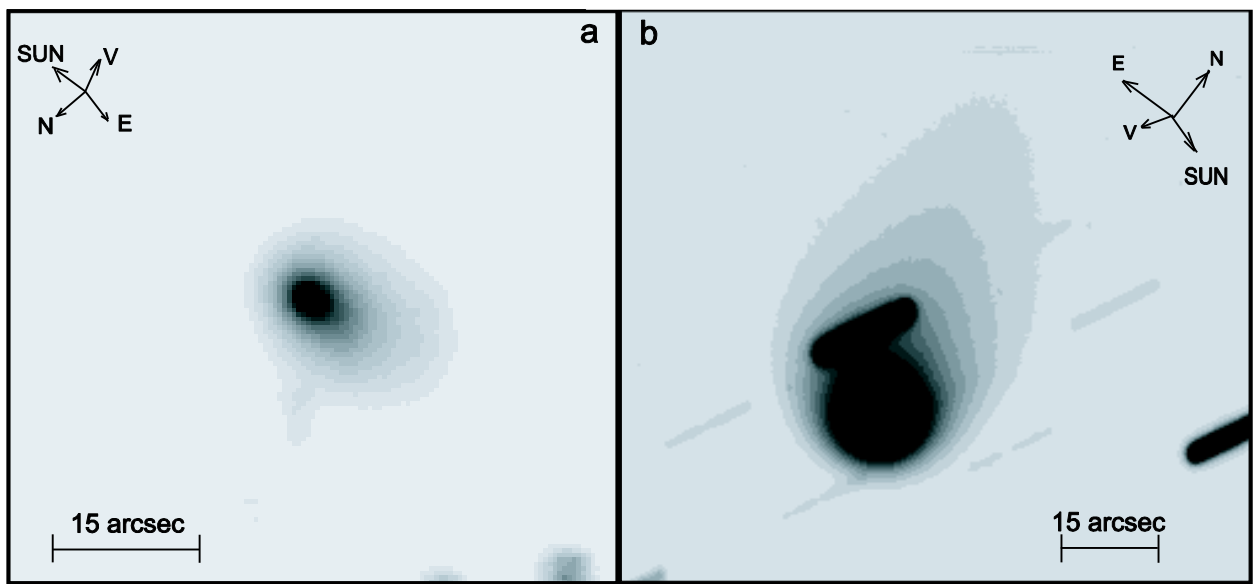

**Fig. 1.** Morphology of comet C/2010 S1 (LINEAR) during two periods of observations: (a) November 25, 2011; (b) November 12, 2012. The arrows show the directions to the Sun, North, East and motion of the comet.

The variations of the intensity along the sunward and tailward directions versus the photocentric distance ($\rho$) in log scale are plotted in Fig. 2 for two dates of observations of comet C/2010 S1. The profiles are dominated by the seeing when the photocentric distance is less than ~8000 km (Fig. 2a) and ~5000 km (Fig. 2b). During the both periods of observations the intensity decreases with the increasing distance from the photocenter. It is noticed that on November 25, 2011 the intensity is slightly higher in the tailward direction as compared to the solar direction. At the projected distance 8000 km ÷ 89000 km along the sunward direction the intensity of coma falls off very rapidly to the sky level with the slope of −1 and the slope is equal to −1.2 along the tailward direction. The radial profiles of the intensity obtained on November 12, 2012, differ considerably from those for the first date. The slope of the intensity increases significantly with increasing distances from the photocenter. Over the range of the cometocentric distances 8000 km ÷ 32000 km, the slope is on average close to − 0.7, and it is equal to − 0.9 for larger distances (up to 80000 km). At the projected distance in excess of 80000 km, the curves show a rapid drop with distance according to $\rho^{-1.28}$ (in the sunward direction) and $\rho^{-1.31}$ (in the tailward direction) dependences, which are close to those derived for the first observation of the comet.

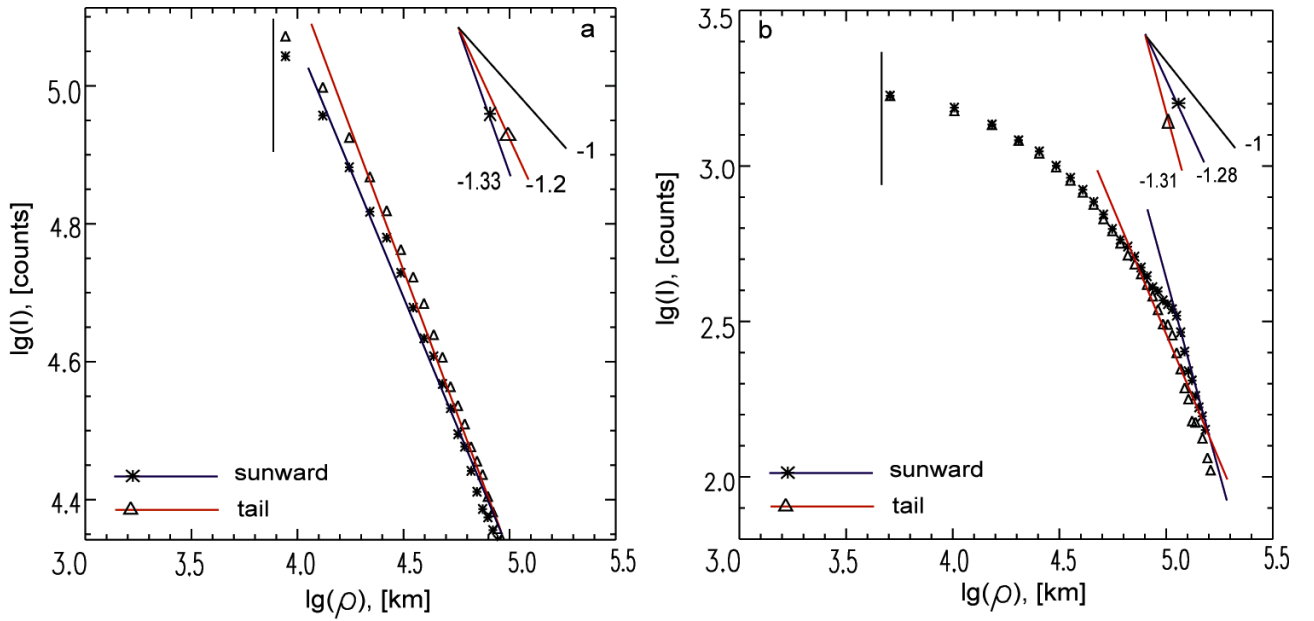

**Fig. 2.** Cuts through the coma in the sunward and tailward directions for the two dates of observation of comet C/2010 S1 (LINEAR): (a) November 25, 2011; (b) November 12, 2012. Vertical lines indicate the seeing radius limit. The radial profile of isotropic coma is represented with a slope of −1 in log scale.

The intensity image of the second observed comet C/2010 R1 and the intensity map with isophotes are shown in Fig. 3 (a and b, respectively). As one can see, this comet had well-concentrated coma and extensive dust tail typical of some distant comets (Reomer, 1962, Korsun and Chorny, 2003, Korsun, 2005 and Korsun et al., 2010).

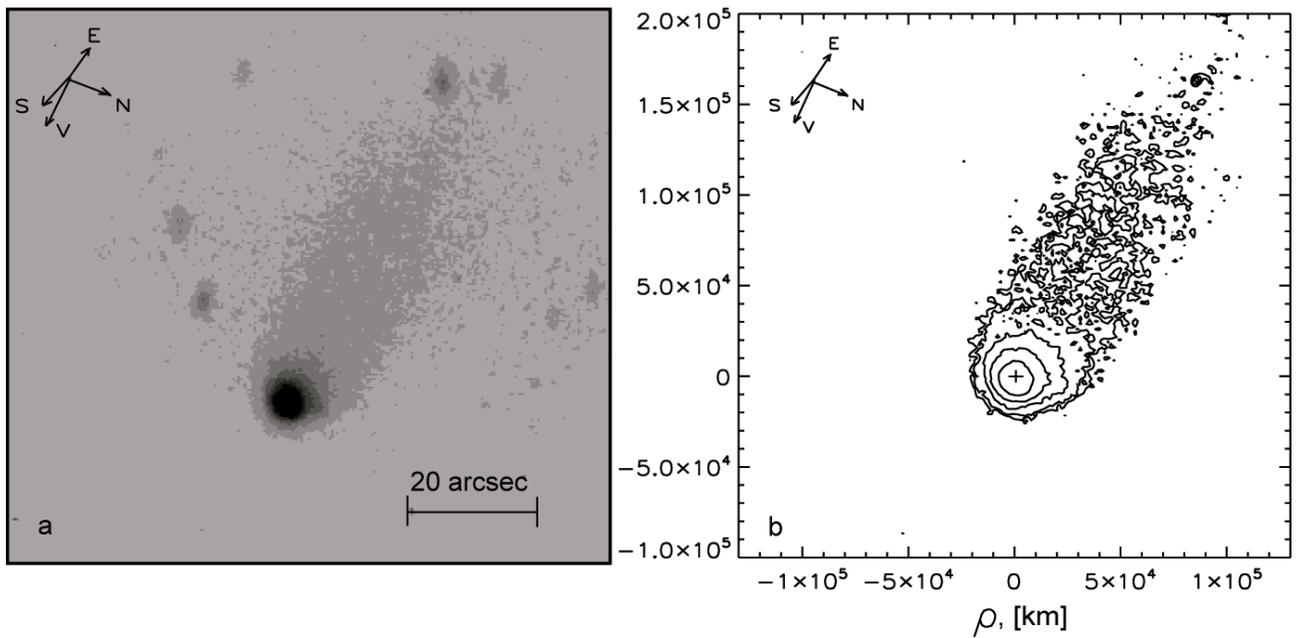

**Fig. 3.** Morphology of comet C/2010 R1 (LINEAR) on February 6, 2013: (a) intensity image obtained through the wide r-sdss filter; (b) intensity map with the isophotes differing by a factor $\sqrt{2}$. The "+" mark denotes the photocenter of the comet. The arrows show the directions to the Sun, North, East and motion of the comet.

The radial profiles of the intensity and their comparison with an isotropic coma are shown in Fig. 4. Since the comet was observed with long tail, we constructed profiles along the sunward and tailward directions. The differences between the sunward and tailward profiles are significant. The intensity of coma is considerably higher in the sunward direction as compared to the tailward direction at distances up to 16000 km from the photocenter, whereas at larger distances the intensity of tail is higher. The slope near the photocenter of the comet is varying from $-0.89$ to $-1.19$ for the sunward and tailward directions, respectively. The decrease in intensity as a function of the photocentric distance is close to $-1.29$ for distances more than 16000 km in the sunward direction and is close to $-1$ in the direction of tail. Since the slope increases progressively with the optocentric distance in the solar direction, it may indicate that the dust is moving from this region. A slope smaller than $-1$ indicates that the dust is pushed in the tail by the solar radiation, replenishing the antisolar regions.

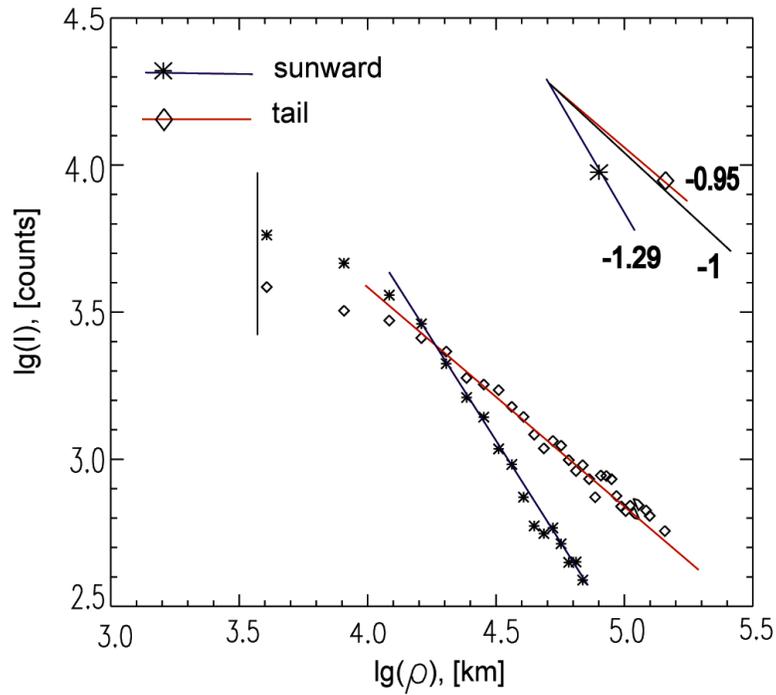

**Fig. 4.** Cuts through the coma of comet C/2010 R1 (LINEAR) in the sunward and tailward directions. Slopes of radial intensity profiles −1.29 and −0.95 are indicated. Isotropic coma represented as −1 line. Vertical line indicates the seeing radius limit.

## 3.2. Polarimetry of comets

### 3.2.1. Comet C/2010 S1 (LINEAR)

Polarization maps of comet C/2010 S1 for two dates of observations are presented in Figs. 5 and 6. The left panels show spatial distribution of the degree of linear polarization over the coma, while the right panels display variations of the polarization along the sunward and perpendicular directions. For both dates, the errors in polarization are almost consistently lower in the inner coma of the comet as compared to the slightly higher value in the outer coma. Polarization is found to vary over the coma in both directions, however, in different ways. On November 25, 2011, the minimum polarization degree (of about − (1.9±0.1)% is noticed in the circumnuclear area of the comet. Polarization increases (in absolute value) gradually up to about −(2.7±0.2)% with an increase of the photocentric distance up to 10000 km in the sunward direction, and it increases up to −(2.5±0.2)% in the tailward direction up (to 25000 km). Such behavior of polarization is also observed in the perpendicular direction (see Fig. 5). On November 12, 2012, the degree of polarization does not practically change, and its value over the coma is within the range from −(1.8±0.2)% up to −(3.0±0.3)% (see Fig. 6). The profile of polarization in the sunward direction practically coincides with the profile in the perpendicular direction.

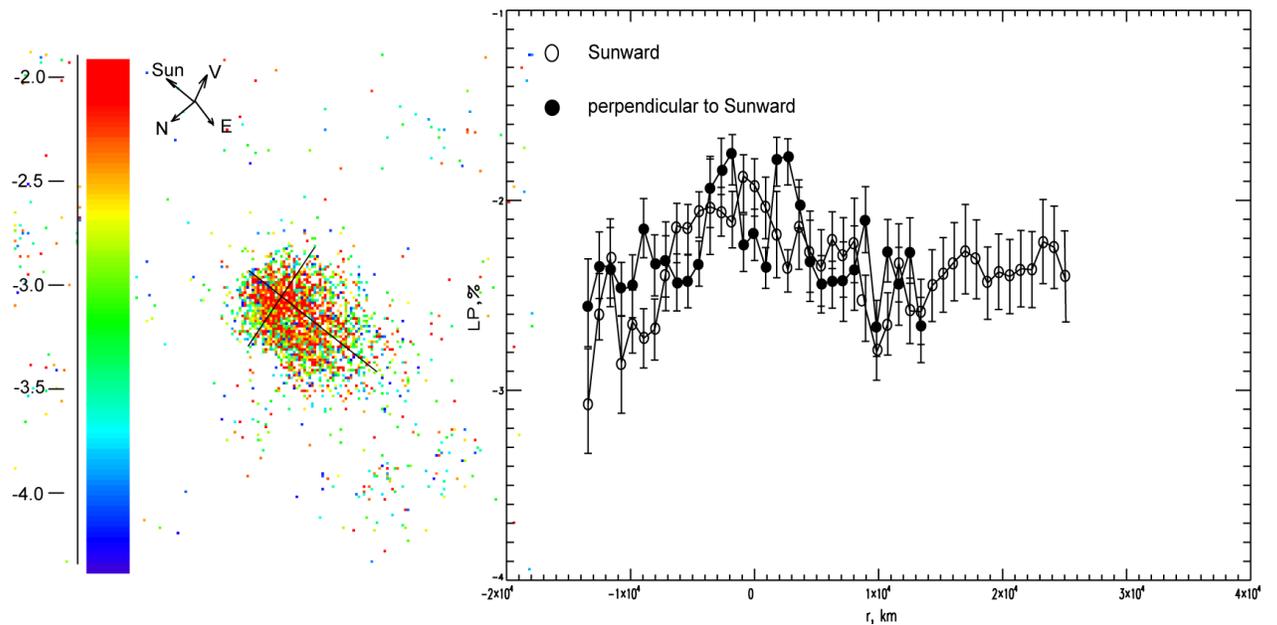

**Fig. 5.** Polarization map and variations of the degree of polarization of comet C/2010 S1 (LINEAR) along the sunward and in perpendicular directions on November 25, 2011. The arrows show the directions to the Sun, North, East, and motion of the comet.

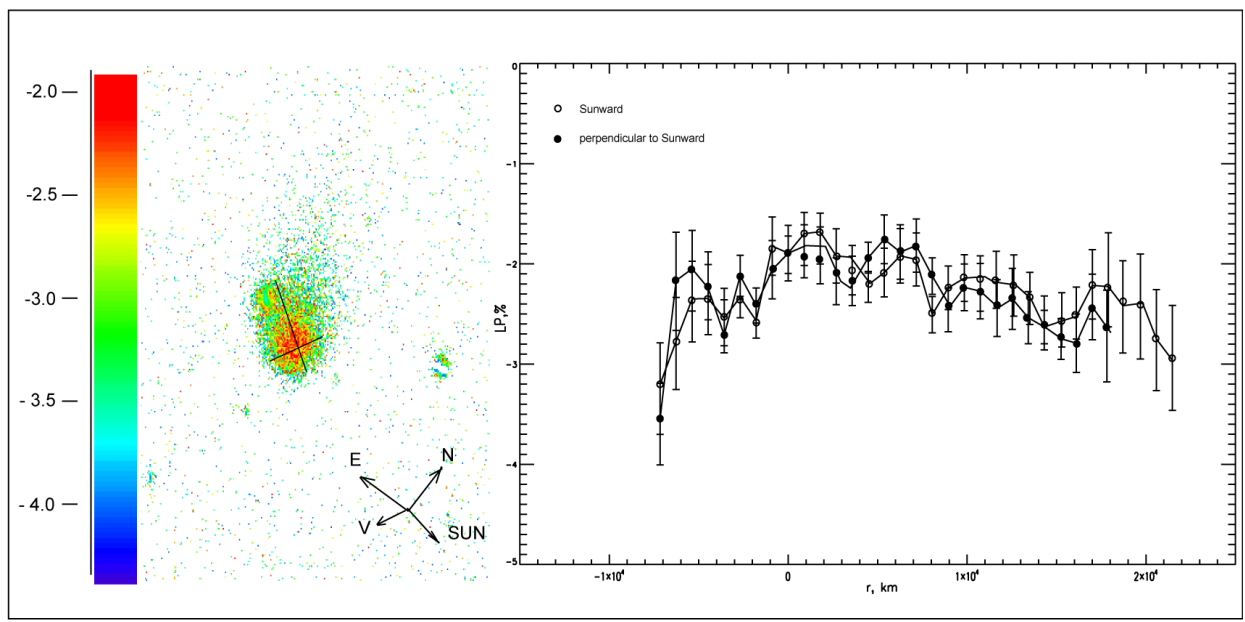

**Fig. 6.** Polarization map and variations of the degree of polarization of comet C/2010 S1 (LINEAR) along the sunward and in perpendicular directions on November 12, 2012. The arrows show the directions to the Sun, North, East, and motion of the comet.

### 3.2.2. *Comet C/2010 R1 (LINEAR)*

Results of observations of comet C/2010 R1 are presented in the form of the polarization map (Fig. 7, left panel) and cuts along the sunward and perpendicular directions (Fig. 7, right panel). The polarization degree of comet C/2010 R1 is generally higher (in absolute value) in comparison with comet C/2010 S1. Polarization profiles in the sunward and perpendicular directions are identical in the inner coma up to the distance from the photocenter equal to 10000 km. We can see that increasing polarization in the cometary coma goes from the center to the periphery (along the tail). The central region of the coma (up to 5000 km) shows polarization of $-(2.50\pm0.05)\%$ which increases almost to $-(3.9\pm0.1)\%$ at the distance of about 22000 km from the photocenter in the tailward direction. Variation of the polarization degree with the distance from the photocenter can be an evidence of differences in the physical properties of the dust over the coma and tail.

Variation of polarization degree with distance from the photocenter can be an evidence of differences in the physical properties of the dust over the coma and tail. Imaging polarimetry of comets revealed dramatic spatial variation of linear polarization throughout coma (Hadamcik and Levasseur-Regourd, 2003; Jewitt, 2004). For instance, at small phase angles (< 30) there were found two polarimetric features, the so-called circumnucleus halo with an extreme amplitude of the negative polarization $P_{min}$ = -6%, and jets revealing only positive polarization (Hadamcik and Levasseur-Regourd, 2003). At side scattering (phase angle ~90 degree), the polarization may also cover a significant range, from ~3% up to more than 30% (Jewitt, 2004). Hadamcik and Levasseur-Regourd (2003) attributed variations of the linear polarization to different physical properties of cometary dust. Namely, there was suggested that the strong negative polarization in the circumnucleus halo is produced by particles with compact morphology; whereas, the posiive polarization in jets at small phase angles results from fluffy morphology of dust particles. This suggestion was examined in Zubko et al. (2008). There was found the polarimetric response near backscattering to be weakly dependent on packing density of irregularly shaped particles. On the contrary the difference in polarization of the halo and jets was interpreted in terms of different material absorption of the constituent particles (Zubko et al., 2009). Later the difference in material absorption was demonstrated to be the most probable reason for spatial variations of polarization observed at side scattering (Zubko et al., 2013). However, Zubko et al. (2013) also noticed an impact of size distribution on polarimetric response at side scattering, which could explain some observations of comets.

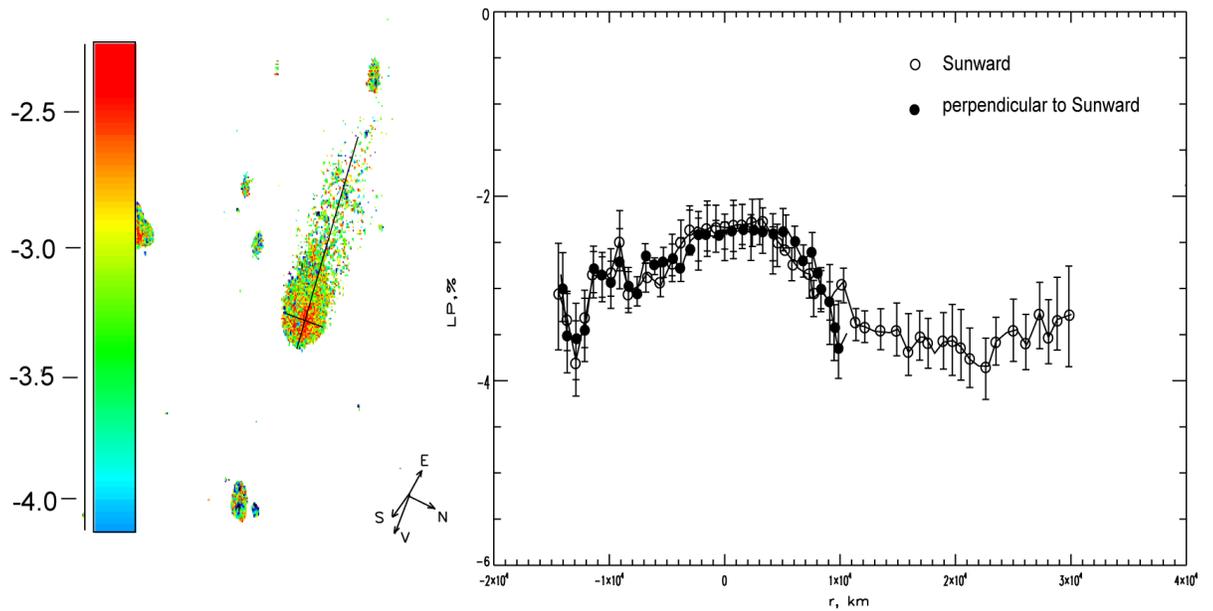

**Fig.7.** Polarization map and variations of the degree of polarization of comet C/2010 R1 (LINEAR) along the sunward and in perpendicular directions on February 6, 2013. The denotations are the same as in Fig. 5.

Figure 8 shows the averaged values of linear polarization for comets C/2010 S1 and C/2010 R1 in comparison with the mean phase-angle dependence of polarization for dusty comets (i.e., comets with high dust-to-gas ratio) (Kiselev and Rosenbush, 2004). Obviously, the polarization degree for these two distant comets reveals significant deviation from the trend shown in Fig. 8 with solid line. The same values of polarization were obtained for only two comets observed at small heliocentric distance. For comet C/2002 T7 (LINEAR) in V filter at phase angle of 6.77 deg, polarization P was found to be –(2.33±0.35)%, whereas P = – (1.99±0.17)% at phase angle of 6.59 deg (Rosenbush et al. 2006). Both values are consistent with the data obtained for comet C/2010 S1 (LINEAR). The value of polarization obtained for comet C/2010 R1 (LINEAR) agrees quantitatively with that measured for comet C/1975 V1 (West) (Kiselev and Chernova (1980)).

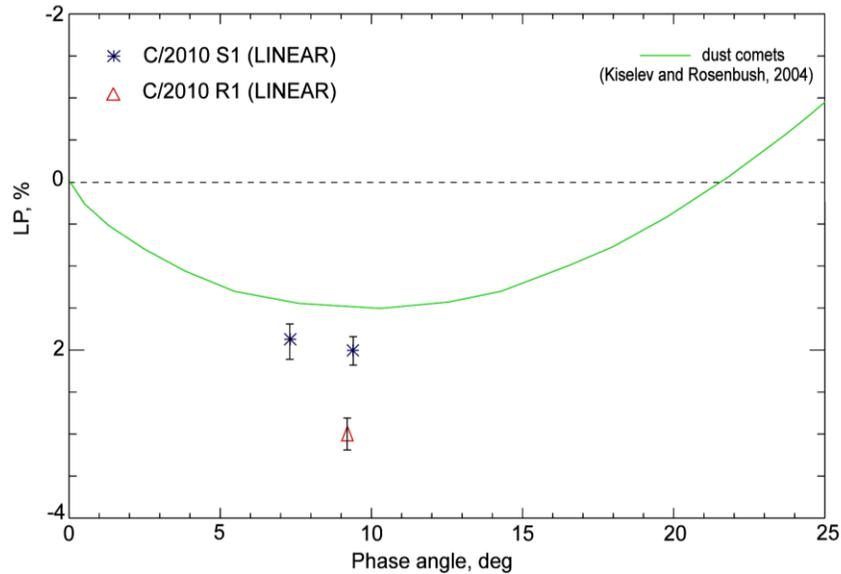

**Fig.8.** Comparison of the whole coma polarization for comets C/2010 S1 (LINEAR) and C/2010 R1 (LINEAR) with the average phase angle dependence of polarization for the dusty comets (solid curve).

## 4. Numerical modeling

Until now, numerical simulations of light scattering characteristics have been performed in order to determine physical characteristics of cometary dust on the basis of available photopolarimetric observational data obtained for comets close to the Sun (e.g. Kolokolova et al., 2003, Petrova et al., 2004, Kimura et al., 2006, Kolokolova et al., 2007, Das et al., 2008, and Zubko et al., 2011, 2014). Even though photopolarimetric data for comets have been obtained within a broad range of phase angles and wavelengths, so far there is no ultimate conclusion on the nature, shape, size of the dust in atmospheres of these comets. Usually when analyzing the results of polarimetric observations, an essential attention is concentrated on the reproduction of the negative branch of linear polarization in the range of small phase angles, and the model of fractal-like aggregates composed of submicron spherical monomers is often used for this purpose (Kolokolova et al., 2003, Petrova et al., 2004, Kimura et al., 2006, Kolokolova et al., 2007, Das et al., 2008, and Rosenbush et al., 2009). At the same time for comets 17P/Holmes and C/1975 V1 (West), it was shown the results of photopolarimetric observations to be consistent with the model of cometary atmosphere composed of high absorbing agglomerated debris particles (Zubko et al., 2011, 2014).

As was anticipated above, our polarimetric data are the first ever obtained for distant comets. From these data, it can be expected that comets C/2010 S1 and C/2010 R1 have much deeper branch of negative polarization as compared with the case of comets close to the Sun. Here, we make an attempt to estimate possible optical properties of the dust in the atmosphere of distant comets by looking for good agreement between the observational data and the results of numerical modeling.

As usually for comets, we suppose a low concentration of dust what enables to consider a cometary atmosphere as an optically thin layer and to exclude the contribution of multiple scattering. For a macroscopically isotropic and mirror-symmetric medium the far-field transformation of the Stokes parameters upon the scattering is written in terms of the normalized Stokes scattering matrix $\mathbf{F}(\theta)$

$$\begin{bmatrix} I^{sca} \\ Q^{sca} \\ U^{sca} \\ V^{sca} \end{bmatrix} \propto \mathbf{F}(\theta) \begin{bmatrix} I^{inc} \\ Q^{inc} \\ U^{inc} \\ V^{inc} \end{bmatrix} \propto \begin{bmatrix} a_1(\theta) & b_1(\theta) & 0 & 0 \\ b_1(\theta) & a_2(\theta) & 0 & 0 \\ 0 & 0 & a_3(\theta) & b_2(\theta) \\ 0 & 0 & -b_2(\theta) & a_4(\theta) \end{bmatrix} \begin{bmatrix} I^{inc} \\ Q^{inc} \\ U^{inc} \\ V^{inc} \end{bmatrix}, \quad (5)$$

where $\theta \in [0°, 180°]$ is the angle between the incidence and scattering directions ($\theta = 180° - \alpha$, $\alpha$ is the phase angle), and both sets of the Stokes parameters are defined with respect to the scattering plane (Mishchenko et al., 2002). If the incidence light is unpolarized, then the element $a_1$ characterizes the angular distribution of the scattered intensity, while the ratio $P = -b_1/a_1$ gives the corresponding degree of linear polarization.

In our numerical modeling, we represent cometary particles in the form of fractal aggregates composed of spherical monomers and described by the following statistical-scaling law (Sorensen, 2001)

$$N = k_0 \left( \frac{R_g}{a} \right)^{D_f}, \quad (6)$$

where $a$ is the monomer radius, $1 \leq D_f \leq 3$ is the fractal dimension, $k_0$ is the fractal prefactor, $N$ is the number of monomers in the aggregate, and $R_g$, called the radius of gyration, is a measure of the overall aggregate radius. Both $D_f$ and $k_0$ specify the morphology of a fractal aggregate. Densely packed aggregates have $D_f$ values close to 3, whereas the fractal dimension of chain-like and branched clusters can be much smaller. The prefactor $k_0$ is also related to the compactness state of a fractal such that, for a fixed $D_f$, the packing density tends to be smaller as $k_0$ decreases (e.g. Liu and Mishchenko, 2005). In order to generate monomer positions in a fractal cluster, we use the simulated diffusion-limited aggregation (DLA) method (Mackowski, 1995) in which the generation procedure starts with a pair of spheres in contact for pre-set $k_0$ and $D_f$ values and adds a single monomer at a time. Figure 9 shows examples of aggregates generated for $D_f = 2.8$, $k_0 = 1.06$ (a), and $D_f = 3$, $k_0 = 1$ (b).

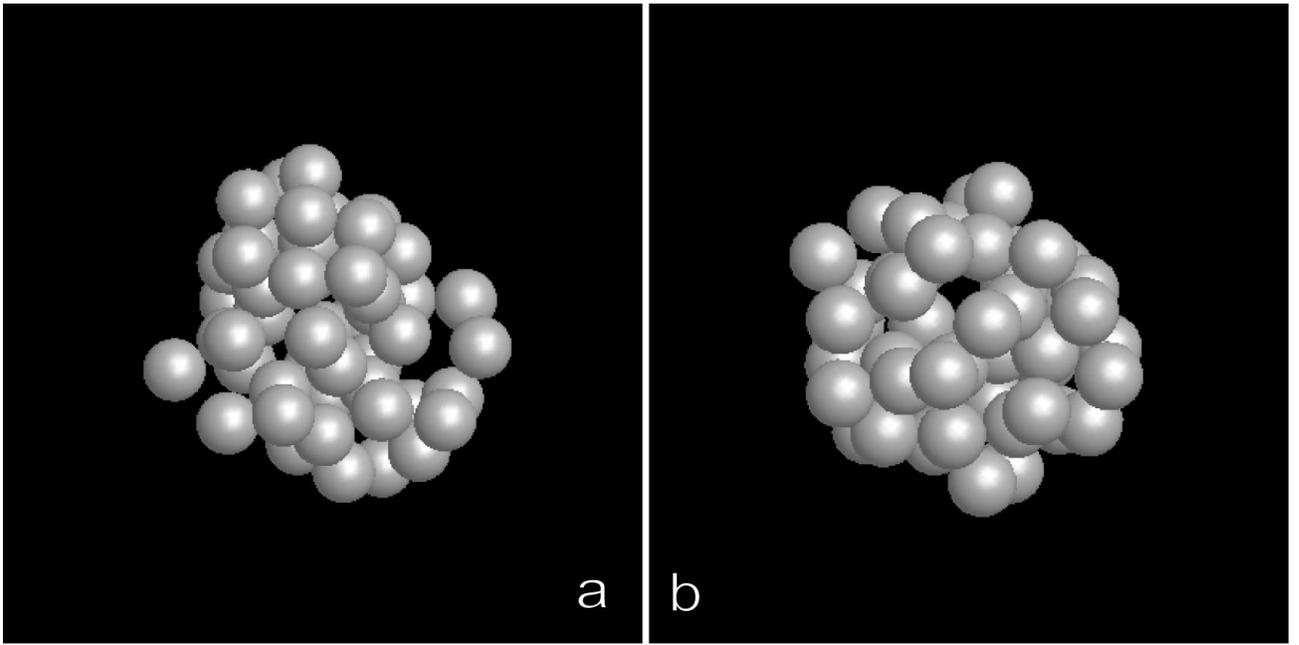

**Fig. 9**. Examples of considered aggregate particles: $D_f = 2.8$, $k_0 = 1.06$ (a), and $D_f = 3$, $k_0 = 1$ (b).

We have performed numerous computations of light scattering characteristics for different sizes and chemical composition of cometary dust by applying the superposition T-matrix method developed for multisphere particle groups in random orientation (Mackowski and Mishchenko, 1996, 2011) and the parallelized computer code (http://www.eng.auburn. edu /users /dmckwski /scatcodes). Here, we show our results obtained for the refractive index $m = m_R + i\,m_I = 1.65 + i\,0.05$. Note that this value of the refractive index can be adopted for cometary dust composed mainly of astronomical silicates (Petrova et al., 2004). Most of computations were performed for wavelength $\lambda = 0.55$ μm, and our main task was to analyze the observational data obtained for comet C/2010 S1.

Figure 10 depicts the computed angular dependences of the degree of linear polarization on the phase angle for medium composed of fractal aggregates for the monomer size parameter $x = 2pa/\lambda = 1.5$ (the corresponding value of the monomer radius $a = 0.13$ μm at $\lambda = 0.55$ μm), the number of monomers $N = 40, 50, 100$, and $D_f = 2.8$, $k_0 = 1.06$ (the upper row), $D_f = 3$, $k_0 = 1$ (the bottom row). It is seen that for both types of aggregates there is no satisfactory agreement between the results of computations and the observational data for comet C/2010 S1.

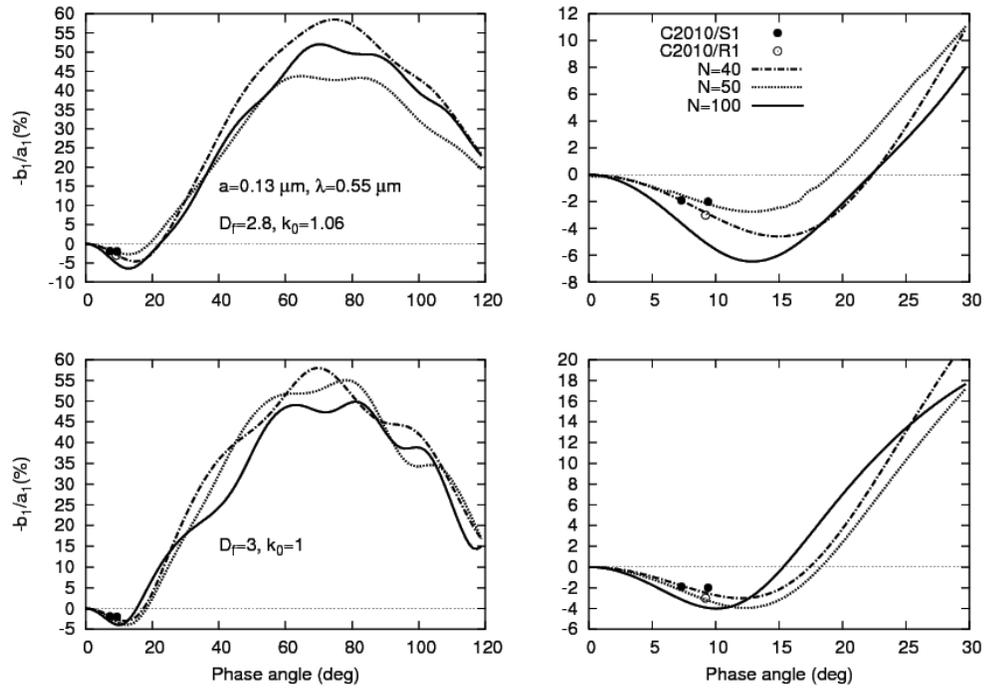

**Fig. 10**. Phase angle dependences of the degree of linear polarization of light scattered by aggregates with $D_f = 2.8$, $k_0 = 1.06$, composed of different number of monomers ($N = 40, 50, 100$), $a = 0.13$ μm at $\lambda = 0.55$ μm (the upper row), and $D_f = 3$, $k_0 = 1$ (the bottom row).

We have also performed similar computations for $D_f = 3$, $k_0 = 1$ and $a = 0.1$ μm ($x = 1.14$). Note that the overall shape of this aggregate is nearly spherical with radius $R \sim 0.6$ μm. The upper row of Fig.11 presents the results obtained for $\lambda = 0.55$ μm. In the case of $N = 100$, the obtained data are consistent to some extent with observations. But the oscillations in the behavior of the degree of polarization (the left panel) are not inherent to comets. In the bottom row, we show the computed phase angle dependences obtained for $N = 100$ and wavelength of $\lambda = 0.45, 0.55$, and $0.9$ μm. It is seen that for $\lambda = 0.9$ μm and small phase angles, the negative branch of polarization vanishes.

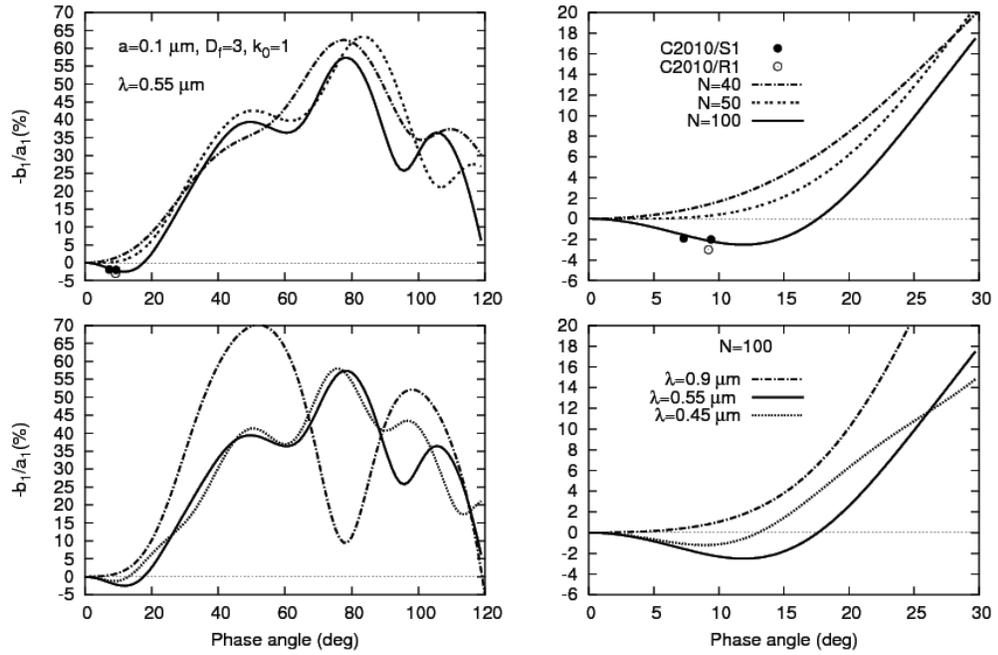

**Fig. 11**. Phase angle dependences of the degree of linear polarization of light scattered by aggregates with $D_f = 3$, $k_0 = 1$, composed of different number of monomers ($N = 40, 50, 100$), $a = 0.1$ μm at $\lambda = 0.55$ μm (the upper row), and $N = 100$ at $\lambda = 0.45, 0.55, 0.9$ μm (the bottom row).

It must be noted that all the results given in Figs. 10 and 11 were obtained by averaging all single scattering characteristics over the ensemble of five fractal-parameter-equal generations of an aggregate. Desirableness of the averaging was motivated, in particular, by Mishchenko and Dlugach (2009).

At the next step of our investigation, we have examined aggregates composed of large number of monomers. It must be noted that numerically exact modeling light scattering characteristics for aggregates composed of large number of monomers became practically possible due to the superposition T-matrix parallelized computer code (Mackowski, http://www.eng.auburn.edu/users/dmckwski/scatcodes). In particular, this code was used by Kolokolova and Mackowski (2012) when analyzing polarization of light scattered by large aggregates. The results of our computations are presented in Fig. 12. The upper row presents the results of computations performed for $D_f = 3$, $k_0 = 1$, $x = 0.5$ ($a = 0.044$ μm for $\lambda = 0.55$ μm), $N = 1000$ and 2000, and $x = 1.14$ ($a = 0.1$ μm for $\lambda = 0.55$ μm), $N = 1000$. It is seen that in all cases the results of computations do not agree with the observational data. Also, the computed phase curves of polarization show strong oscillations, not typical of comets.

Besides, we considered the case of aggregate particle with $D_f = 2.8$, $k_0 = 1.06$, $a = 0.1$ μm ($x = 1.14$), and $N = 1000$. The overall radius of this aggregate $R$ is of about 1.3 μm what corresponds to size parameter $X \sim 15$ at $\lambda = 0.55$ μm, and the packing density of monomers $c = N(a/R)^3 \sim 0.46$

(porosity $p \sim 0.54$). The results of computations are depicted in the bottom row of Fig. 12. One can see that for comet C/2010 S1, the model of cometary dust composed of aggregates with such fractal parameters allows to obtain sufficiently good agreement between the results of observations and computations. Also, we present the computed curves of the degree of linear polarization for wavelengths $\lambda = 0.45$ and $0.9$ $\mu$m (the corresponding values of the size parameter of monomers are equal to 1.4 and 0.7). One can see that for all wavelengths the negative branch of polarization exists in the range of small phase angles. It should be noted that the dependence of the depth of the negative branch is not monotonic function on the wavelength (or on the size parameter of monomer $x$). Also, this conclusion is confirmed by the data obtained for $N = 100$ (see Fig. 12, the bottom row). Moreover, the dependence on the number of $N$ is nonmonotonic also (compare Figs. 11 and 12, right panel in the upper rows). This result contradicts the conclusions obtained, for instance, by Petrova et al. (2004) and Kimura et al. (2006). Such behavior of the negative branch of polarization is indicative of the complexity of the physical mechanisms involved in its formation.

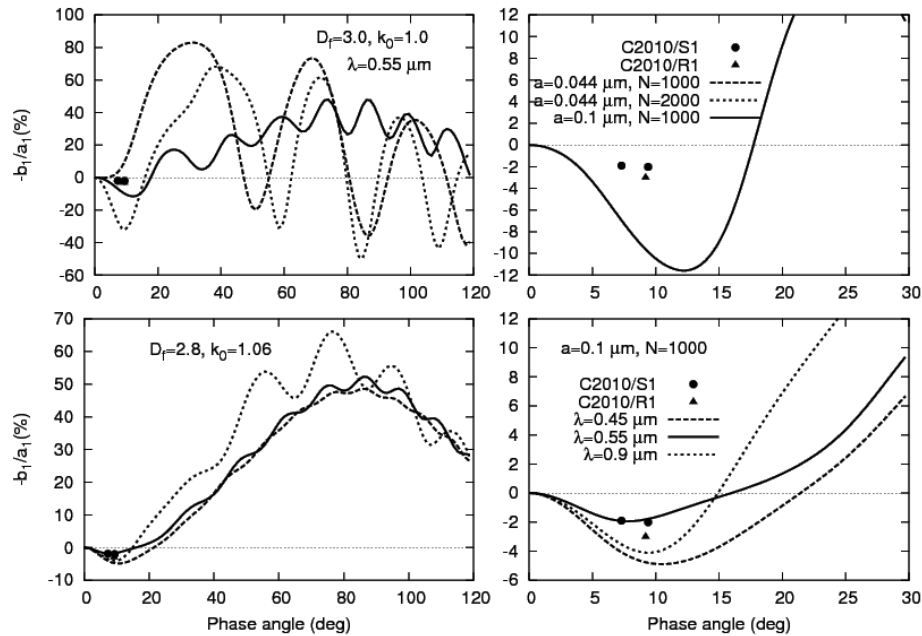

**Fig. 12.** As in Fig. 11, but for $D_f = 3$, $k_0 =1$, (the upper row), and $D_f = 2.8$, $k_0 =1.06$, (the bottom row).

Figure 13 depicts the corresponding computed phase angle dependences of the scattered intensity (the single scattering matrix element $a_1$) for $\lambda = 0.45$, 0.55, and 0.9 $\mu$m. It is seen that the backscattering peak increases with decreasing the wavelength (or increasing the monomer size parameter). Also in the range of large phase angles, the oscillations in the behavior of the intensity (element $a_1$) are observed. It should be noted that such behavior of polarization and intensity (obtained from computations) is typical of spherical particles, and it was not observed for the comets close to the Sun. It is possible that these oscillations can be smoothed out when averaging over different sized aggregates as well as by polydispersion of different sized monomers. But it should be noted that this averaging would also result in changes in the behavior of the negative branch of polarization in the range of small phase angles. In our opinion, the effect of sized polydispersion for aggregates, composed of large number of submicron spherical monomers, requires a special non-trivial research.

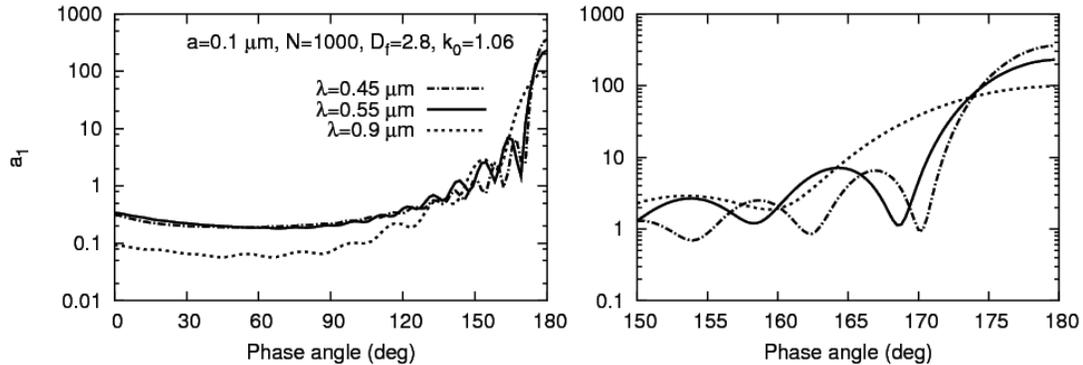

**Fig. 13**. Phase angle dependences of the scattered intensity by aggregates composed of $N = 1000$ monomers with $a = 0.1$ μm at $\lambda = 0.45, 0.55, 0.9$ μm.

5. **Discussion**

The main aim of our observations of distant comets is to study their dust properties. Polarimetric observations of distant active comets with specific features (such as long tails, jets structures, and asymmetric comae) are of particular interest. Our photometric and polarimetric observations of two dynamically new comets C/2010 S1 (LINEAR) and C/2010 R1 (LINEAR) showed a considerable level of activity (extended dust comae and tails) at heliocentric distances of 5.9 – 7.0 AU. Since molecular emissions were not detected in the spectra of these comets, it allows one to study reflective dust properties without gas emission effect.

As it is known, the shape of the cometary coma as well as the distribution of the surface brightness tell on the cometary activity, but still the detailed mechanisms of the activity of comets at large heliocentric distances are not well understood (Meech et al., 2009). The surface brightness of the coma decreases as a function of the distance $\rho$ from the nucleus according to a power law $I=k\rho^{-n}$, where $k$ is the scaling factor for the coma and $n$ is the slope. In the case when radial dust ejection from the nucleus occurs with a constant rate and with constant speed (a steady-state dust outflow), the radial gradient profile exhibits a constant slope of $-1$ (A'Hearn et al., 1984 and Gehrz and Ney 1992). Even if the dust is emitted radially from only a part of the surface, then a $1/\rho$-distribution of the surface brightness takes place.

The differences between the sunward and tailward intensity profiles are found from our observations of comets C/2010 S1 (LINEAR) and C/2010 R1 (LINEAR). The intensity of coma falls off very rapidly to the sky level. With increasing the distance from the photocenter, the slope of radial profiles of intensity changes gradually from − 0.7 (near the nucleus) to of about − 1.3 (for larger distances up to 100000 km). There are a number of different processes which can cause the profile to steepen or flatten out. The deviations of the slope of intensity from the canonical value of −1 may indicate the existence of changes in the dust, which are generally caused by some reasons (i.e., the temporal changes in the dust production, the effects of radiation pressure (segregation of the particles according to their size and mass), some variation in the physical properties of the grains (fading, fragmenting, sublimating) etc) (Jewit and Meech, 1987, Jewit and Luu, 1989, Boice et al., 2000 and Lamy et al., 2009). However, all these processes typical of comets at small heliocentric distances also can be observed for comets at heliocentric distances up to 5 AU. The sublimation of water ice is the main driver of activity for comets close to the Sun. But even in this case, an extremely high fragmentation rate would be required to account for the observed deviations from a $\rho^{-1}$-dependence in the intensity profiles (Konno et al., 1992). Thus, some other mechanisms must be invoked to explain comet activity at large heliocentric distances.

Due to the low surface temperature, sublimation of pure water ice is inefficient at distances larger than 3 AU. However, Meech and Svoren (2004) showed that even up to $r$ = 5–6 AU there is a gas flux caused by sublimation, which is sufficient to lift off small grains (<0.1 μm) from the surface. Actually, as Hanner (1981) showed that even slightly dirty ice grains drastically decrease the albedo and increase the grain temperature. At large heliocentric distances, where sublimation rate of icy grains is very low, water ice grains may also be ejected during sublimation of carbon monoxide or other volatile species (Grün et al., 1993). However, the detailed mechanisms of this activity are not well understood, especially, for explanation an unexpected and irregular activity exhibited by many comets at large distances from the Sun. Amorphous-crystalline phase transition of water ice seems to be one more mechanism, which may supply enough energy in order to eject near-surface material and explain appearance of coma at large heliocentric distances what is onsistent with the results of observations (Jewitt, 2009 and Meech et al., 2009). These processes can explain the relatively high activity in new comets and non-isotropic coma at large heliocentric distances and, hence, deviations of the slope of intensity from the canonical value of −1.

The characteristics of polarization can be used in order to estimate size, composition, and structure of dust particles. Previous studies (e.g. Bastien et al., 1986, Kikuchi et al., 1987 and Kiselev and Rosenbush, 2004) have established that polarization of comets decreases with increasing the size of diaphragm. One of possible reasons for such behavior may be gas contamination increasing with size of aperture and lower polarization of light scattered by molecules as compared with that scattered by dust particles. At large phase angles, such trend is typical of comets which spectra usually consist of continuum and molecular emissions.

At phase angles less than 20°, Hadamcik and Levasseur-Regourd (2003) have found inhomogeneous distribution of polarization within the cometary coma, namely, a positive polarization in jets and high negative polarization degree (up to −6%) in the circumnucleus halo. For some comets (but not all), radial variations of polarization are seen in the innermost coma, where the polarization is generally low near the nucleus and increases with distance (Dollfus et al., 1988, Jewitt, 2004 and Kolokolova et al., 2004). This supports the conclusion that grains are fragmenting into smaller particles as they recede from the nucleus. Some structures (e.g., jets) show larger values of polarization than coma (even for the same comet) (Tozzi et al., 1997). This fact suggests that there are fundamental differences in the size and structure of the dust in the jet outflow what can be an indicator of inhomogeneities in nucleus (Tozzi et al., 1997 and Farnham, 2008). However, all these features of polarization are typical of comets near the Sun at the heliocentric distances up to 4 AU. But it should be emphasized that not all comets show such trend. For example, the polarization of comet C/1983 H1 (IRAS−Araki−Alcock) remained high (~30%) in the near-nuclear region at distances ~350−850 km (Kiselev et al., 2006).

The distributions of polarization observed for comets C/2010 S1 and C/2010 R1 showed that the lowest degree of polarization (in absolute value) was around the photocenter of the coma, and it increased slowly with increasing the distance from the nucleus. The possible reasons of changing polarization over the coma and tail may be the same processes which cause an appearance of active coma at large heliocentric distances and subsequent evolution of dust particles properties with increasing the distance from the nucleus. As we have shown, the polarization of comets C/2010 S1 and C/2010 R1 is significantly higher (variations over the coma are from of about −1.9% up to −3.5%) in absolute value than typical value of polarization (~ −1.5%) observed for the whole coma for comets close to the Sun. In particular, it may be caused by the difference in the average characteristics of cometary dust particles for the distant and close to the Sun comets. For example, one cannot exclude that at large heliocentric distances the average size of the grains exported from the surface of the nucleus will be smaller.

The results of our extensive numerical modeling performed by using the superposition T-matrix method show that for comet C/2010 S1 the model of cometary dust in the form of aggregates of overall radius approximately equal to 1.3 μm, composed of 1000 spherical monomers with radius 0.1 μm, porosity $p \approx 0.54$, refractive index $1.65 + i\,0.05$ allows not only to reproduce the negative branch of polarization at small phase angles, but also to obtain a satisfactory quantitative agreement between the results of polarimetric observations and computations. However, we are fully aware that an extremely small number of the available observational data makes quite problematic unambiguous correct determination of the characteristics of the dust particles. Therefore, new photometric and polarimetric observations of distant comets performed in a wide range of wavelengths and phase angles are required to provide a reliable physical model of the cometary dust.

## 6. Conclusions

Our analysis of the results of polarimetric and photometric observations of comets C/2010 S1 (LINEAR) and C/20910 R1 (LINEAR) allows to make the first conclusions on polarization properties of the distant comets:

- Comets show significant activity (the existence of extended coma with long tail) beyond the Jovian orbit. As it is generally observed, the differences between the sunward and tailward profiles are significant. On November 25, 2011 at the projected distance 8000 km ÷ 89000 km, the intensity of the coma of comet C/2010 S1 (LINEAR) falls off very rapidly to the sky level with the slope of −1 along the sunward direction, and the slope is equal to −1.2 along the tailward direction. But the observations performed on November 12, 2012 showed that in the range of cometocentric distances 8000 km ÷ 32000 km, the slope is close to − 0.7, and it is close to − 0.9 for larger distances (up to 80000 km).

- At distances up to 16000 km from the photocenter, the intensity of the coma of comet C/2010 R1 (LINEAR) is considerably higher in the sunward direction as compared to the tailward direction. Near the photocenter, the slope of the comet varies from − 0.89 up to − 1.19 for the sunward and tailward directions, respectively. The decrease in the intensity slope as a function of the photocentric distance is close to − 1.29 for distances more than 16000 km in the sunward direction, and it is close to − 1 in the direction of the tail.

- Polarization maps of comets C/2010 S1 and C/20910 R1 show spatial variations of the polarization. The obtained values of the degree of polarization are significantly higher in absolute values (the variations over the coma are from of about −1.9% up to −3.5%) than the typical value of polarization (~ −1.5%) observed for the whole coma of most comets close to the Sun.

- The results of our numerical modeling show that for comet C/2010 S1 (LINEAR), the cometary dust in the form of aggregates of overall radius $R \sim 1.3$ μm, composed of $N = 1000$ spherical monomers with radius $a = 0.1$ μm, porosity $p \sim 0.54$, refractive index $m = 1.65 + i\, 0.05$, allows to obtain a satisfactory quantitative agreement between the results of polarimetric observations and computations.

## Acknowledgments

We are grateful to Antti Penttilä and anonymous reviewers for useful discussion and comments. The observations were performed due to the support of the Schedule Committee for Large Telescopes (Russian Federation). We thank Dr. V.K. Rosenbush and

Dr. N.N. Kiselev for discussion and constructive criticism of the results of observations. We acknowledge support from the National Academy of Sciences of Ukraine under the Main Astronomical Observatory GRAPE/GPU/GRID Computing Cluster Project.